\documentclass[12pt,preprint,preprintnumbers,amsmath,amssymb]{article}

\usepackage{graphicx}
\usepackage{dcolumn}
\usepackage{bm}

\def\lsim{\mathrel{\lower2.5pt\vbox{\lineskip=0pt\baselineskip=0pt
\hbox{$<$}\hbox{$\sim$}}}}
\def\gsim{\mathrel{\lower2.5pt\vbox{\lineskip=0pt\baselineskip=0pt
\hbox{$>$}\hbox{$\sim$}}}}

\newcommand{\nn}{{\nonumber}}

\newcommand{\gev}{{\mbox{ GeV}\,}}
\newcommand{\mev}{{\mbox{ MeV}\,}}
\newcommand{\be}{\begin{equation}}
\newcommand{\ee}{\end{equation}}
\newcommand{\ba}{\begin{eqnarray}}
\newcommand{\ea}{\end{eqnarray}}

\newcommand{\PRL}[1]{Phys.\ Rev.\ Lett.\ {#1}}

\begin{document}


\begin{center}
{\bf \Large Nature of the axial-vector mesons from their
$N_c$ behavior within the chiral unitary approach}

\vskip .3cm

{\large L. S. Geng, E. Oset}

{\it Departamento de F{\'\i}sica Te{\'o}rica e IFIC,
Centro Mixto Universidad de Valencia-CSIC\\
Institutos de Investigaci\'on de Paterna, Apdo 22085, 46071, Valencia
\ Spain}

\vskip .3cm

{\large J. R. Pel\'aez }

{\it Departamento de F{\'\i}sica Te{\'o}rica II,
 Universidad Complutense de Madrid, 28040   Madrid,\ Spain}

\vskip .3cm

{\large L. Roca}

{\it Departamento de F{\'\i}sica,
 Universidad de Murcia, E-30071,   Murcia,\ Spain}

\end{center}

\begin{abstract}
By describing within the chiral unitary approach the s-wave
interaction of the vector meson nonet with the octet of pseudoscalar
Goldstone Bosons, we find that the main component 
of the axial vector mesons $b_1(1235$),
$h_1(1170)$, $h_1(1380)$, $a_1(1260)$, $f_1(1285)$ and the two states
associated to the $K_1(1270)$ does not follow the QCD dependence on the
number of colors for ordinary $q\bar q$ mesons.
\end{abstract}


\vskip 0.5cm

Even though QCD is well established as the theory of strong
interactions, the fact that it becomes non perturbative in the
hadronic regime makes it very complicated to use within the realm of
hadron spectroscopy. Many states are easily accommodated
within lattice calculations or QCD inspired quark models, but such
calculations in terms of fundamental degrees of freedom, i.e.,
quarks and gluons, are usually troubled with chiral symmetry
breaking, physical masses of quarks or Goldstone Bosons and
their decay widths. In contrast, many models based on
hadronic degrees of freedom cannot extract the quark and gluon
content of hadrons without assumptions hard to relate or justify
within QCD, and frequently have the composition already built in a
priori. Furthermore, all these approaches are complicated by the
possible mixture of states with a different nature.

Most of these caveats can be overcome by studying \cite{Pelaez:2003dy,Pelaez:2006nj}
(\cite{Pelaez:2004xp} for a review) the
dependence on the number of colors, $N_c$, of the poles associated to resonances that appear
in the unitarized meson-meson scattering
amplitudes obtained within a Chiral Effective Theory. See also 
refs.~\cite{GarciaRecio:2006wb,Hyodo:2007np,Roca:2008kr} for works in
the meson-baryon sector.

The relevance of the $1/N_c$ expansion \cite{'tHooft:1973jz} is that
it provides an analytic approach to QCD in the whole energy region
and a {\it clear identification of $q \bar q$ states}, that become
bound as $N_c\rightarrow\infty$ and whose masses scale as $O(1)$ and
their widths as $O(1/N_c)$, without the need for the definition of
valence quarks, or QCD inspired potentials. Other hadronic states
may show different behavior \cite{Jaffe}.

The use of an Effective Theory ensures that all degrees of freedom
below a certain scale are included consistently with the QCD
symmetries. In this respect, chiral symmetry becomes essential,
since it is possible to identify the pions, kaons and the eta with
the eight Goldstone Bosons (GB) associated with the spontaneous
chiral symmetry breaking that is known to exist in QCD, which are
separated by a mass gap of the order of $4\pi f_\pi\simeq 1.2
\,\gev$ from all other hadrons. Actually, since light quarks have a
tiny mass that breaks chiral symmetry explicitly, the lightest
pseudoscalars have a small mass and the mass gap is slightly
reduced. Usually $\Lambda\simeq 1 \,\gev$ is taken as the cutoff of
the QCD Effective Theory, such that $p/\Lambda$, where $p$ is a
typical momentum in the theory, is smaller than one. Since these pseudoscalars are GB, not
only their self-interactions, but the interaction terms with other
fields allowed by chiral symmetry are much reduced. When a well
defined power counting exists, it is possible to make a derivative
and mass expansion in the Effective Lagrangian and calculate loop
corrections whose divergences are absorbed in higher order
parameters, that encode the information on Physics beyond the cutoff
scale. This is the case of Chiral Perturbation Theory (ChPT)
\cite{chpt1}. From QCD it is not possible to calculate the Effective
Lagrangian parameters, but at least it is possible 
to know their dependence on certain QCD
parameters.
One example of relevance for this work is the leading $N_c$ scaling of
the parameters appearing in the effective Lagrangian.

Over the last few years, it has been shown that it is possible to
generate heavier resonances not initially present in the Chiral
Effective Lagrangian by imposing unitarity on two body scattering
amplitudes
\cite{Truong:1988zp,Oller:1997ng,Guerrero:1998ei,Oller:1997ti,Oller:1998zr,
Pelaez:2004xp,Kaiser:1995cy}. The advantage of this approach for
spectroscopy is that these resonances are generated from first
principles, namely chiral symmetry and unitarity, without any
assumption about their existence or their spectroscopic nature.
Finally, since the leading QCD $N_c$ scaling of the Effective parameters is
known, it is possible to obtain the leading $N_c$ scaling of the mass and
width of each generated resonance and check whether or not their leading
$N_c$ behavior corresponds to that predicted by QCD for a $q\bar{q}$
state. Note that we are interested in the large $N_c$ expansion
close to the physical value of $N_c=3$, but not in the $N_c\rightarrow\infty$ 
limit. This limit is interesting on its own, but
the phenomenology of many hadrons is not well described in the limit or for too large values of $N_c$. 
For instance, baryons become infinitely heavy and  mesons become strictly bound, like
the rho meson that becomes a bound state
decoupled from pions as $N_c\rightarrow\infty$, which has little to do with the 
familiar physical rho behavior. Nevertheless, the rho and its contributions
to the effective Lagrangian are very well described
by the $1/N_c$ expansion evaluated at $N_c=3$ \cite{Pelaez:2003dy}.
Similarly, different components within a mixed state could change
their proportions for very large $N_c$. Since we are interested in the nature of physical states, without altering radically their composition,
we will extract their leading $1/N_c$ behavior by studying their $N_c$ dependence not very far from $N_c=3$.

Paradigmatic examples are the lightest scalar and vector octets,
generated with the coupled channel Inverse Amplitude Method (IAM)
using one-loop ChPT meson-meson
amplitudes~\cite{Guerrero:1998ei,Pelaez:2004xp}, which describe
data up to $\sqrt{s}\simeq1.2\,$GeV. Remarkably, light vectors follow nicely a
$q\bar{q}$ behavior, whereas the $N_c$ behavior of the light scalars
is at odds with a predominant $q\bar{q}$ nature \cite{Pelaez:2003dy}. This result has
been confirmed \cite{Pelaez:2006nj} at two loops for the $\rho$ and
$f_0(600)$ mesons, even getting a hint for the latter of a subdominant $q\bar{q}$ component, rising around 1
GeV, most likely due to mixing between light non-$q\bar{q}$ and
heavier $q\bar{q}$ states.
Scalar states with non-$q\bar{q}$ behavior can be obtained, for instance,
from different combination of tetraquarks, \cite{Jaffe}, including, of course,
the ``molecular'',  or $\pi\pi$ resonance, rearrangement.

In this work we study the QCD leading $N_c$ behavior of axial vector mesons.
In this case there is no Effective Theory available with higher order terms
with a clear chiral counting and $N_c$ behavior. Thus, we are working under
the assumption, already shown to work remarkably well
\cite{Lutz:2003fm,Roca:2005nm,Geng:2006yb}, that they are predominantly
dynamically generated states, and can be generated by a coupled channel
unitarization of an Effective chiral Lagrangian
describing the interaction between light vectors (V) and the pseudo GB octet
(P). This meson-meson state interpretation comes out naturally as long as 
the cutoff in our model has a very natural size in terms 
of $f_\pi$ and meson physics, although some tetraquark arrangements have similar $N_c$
behavior and cannot be excluded.

In brief, the formalism of \cite{Roca:2005nm} uses the standard 
construction of non-linear Effective Lagrangians
\cite{Coleman:1969sm}
to build a chiral Lagrangian with the lowest number of derivatives that follows, which, properly normalized reads
\cite{Birse:1996hd}: \be\label{eq:lag} {\cal
L}=-\frac{1}{4}\{(\nabla_\mu V_\nu-\nabla_\nu V_\mu)(\nabla^\mu
V^\nu-\nabla^\mu V^\nu) \}, 
\quad 
\ee 
where $\nabla_\mu
V_\nu=\partial_\mu V_\nu+[\Gamma_\mu,V_\nu]$ is the covariant
derivative SU(3) matrix with the SU(3) connection defined as
$\Gamma_\mu=(u^\dagger \partial_\mu u+u\partial_\mu u^\dagger)/2$,
$u=\exp(P/\sqrt{2} f_\pi)$ and
\begin{eqnarray}
&&P\equiv\left(
\begin{array}{ccc}
\frac{\pi^0}{\sqrt{2}}+\frac{\eta_8}{\sqrt{6}}& \pi^+& K^+ \\
\pi^-&-\frac{\pi^0}{\sqrt{2}}+\frac{\eta_8}{\sqrt{6}}& K^0\\
K^-& \bar{K}^0&-\frac{2\eta_8}{\sqrt{6}}
\end{array}
\right) ,
\;\\ \nonumber
&&V_\mu\equiv\left(
\begin{array}{ccc}
\frac{\rho^0}{\sqrt{2}}+\frac{\omega}{\sqrt{2}}& \rho^+& K^{*+} \\
\rho^-&-\frac{\rho^0}{\sqrt{2}}+\frac{\omega}{\sqrt{2}}& K^{*0}\\
K^{*-}& \bar{K}^0&\phi
\end{array}
\right)_\mu.
\end{eqnarray}
Note that ideal $\phi-\omega$ mixing has been assumed and that the neglected $\eta'$ effects
could be included at higher orders.

Expanding the Lagrangian of Eq.~(\ref{eq:lag}) to two vectors and
two pseudoscalars one obtains the simple result of
\be\label{eq:lag2} 
{\cal L}=-\frac{1}{4f_\pi^2}\langle[V^\mu,\partial^\nu V_\mu]
[P,\partial_\nu P]\rangle, \quad 
\ee
which has been used in 
refs.~\cite{Lutz:2003fm,Roca:2005nm}.

The $\mathrm{VP}\rightarrow \mathrm{V'P'}$ amplitudes
\cite{Roca:2005nm} are now easily obtained and their dynamics
depends on just one parameter, the pion decay constant $f_\pi$. The
relevant remark here is that QCD fixes $f_\pi\sim O(\sqrt{N_c})$
whereas the V and P masses behave as $O(1)$.

In ref.\cite{Roca:2005nm}, the unitarization of the tree level
T-matrix was carried out within a coupled channel Bethe-Salpeter
formalism for the two meson states: 
\be\label{eq:tree} T=-[1+V \hat
G]^{-1}V\,\vec \epsilon \cdot \vec \epsilon', 
\ee
 with $\vec\epsilon, \vec \epsilon'$ the V and V' polarization vectors, and $V$
of Eq.~(\ref{eq:tree}) is the s-wave projected
 scattering amplitude for the
vector mesons with pseudoscalar mesons obtained from
Eq.~(\ref{eq:lag}):

\be
V_{ij}(s)=-\frac{\epsilon\cdot\epsilon'}{8f^2} C_{ij}
\left[3s-(M^2+m^2+M'^2+m'^2)
-\frac{1}{s}(M^2-m^2)(M'^2-m'^2)\right].
\label{eq:Vij}
\ee
where
$M(M')$,  $m(m')$ correspond to the initial(final) vector mesons
and initial(final) pseudoscalar mesons respectively.
 The indices $i$ and $j$
represent the initial and final $VP$ states respectively and the 
$C_{ij}$ coefficients are given in ref.~\cite{Roca:2005nm}.

In ref.~\cite{Roca:2005nm} a separation is made in the longitudinal and
transverse parts of the amplitude and the poles are shown to appear in the
transverse part, which is the one shown in Eq.~(\ref{eq:tree}).

As shown in ref.\cite{Roca:2005nm}, the matrix $\hat G$ is diagonal with the $l-$th
element $\hat G_l=(1+{1\over 3} {q_l^2\over M_l^2})\,G_l$ where the
 term with the on-shell center of mass momentum of the intermediate
  states $q_l$ amounts to a few percent,
and $G_l$ is
the two-meson loop  function:
\be G_l(P)=\int\frac{i\, d^4
q}{(2\pi)^4}\frac{1}{(P-q)^2-M_l^2+i\epsilon}
\frac{1}{q^2-m_l^2+i\epsilon}
\label{eq:loopVP}
\ee
where $P$ is the total four-momentum, $P^2=s$. In order to improve
slightly the data description, in \cite{Roca:2005nm,Geng:2006yb} the
finite widths of vector mesons were included in their propagators.
 This correction plays a secondary role in our context, since
these vectors are firmly established $q\bar{q}$ states, and their
widths behave as $1/N_c$.

This Bethe-Salpeter approach is widely used in the
literature~\cite{Kaiser:1995cy}, and is related to other
unitarization techniques as the IAM
\cite{Truong:1988zp,Guerrero:1998ei,Pelaez:2004xp}
or the N/D \cite{Oller:1998zr}, which differ from the one described
above in that they include higher orders, a left cut, or tadpole and
crossed channels. These differences may be relevant at very low
energies but at higher energies,
where we are generating dynamically the present resonances from the lowest
order chiral Lagrangians, they have been shown to be minute.

Since the $G_l$ integral above is divergent, it has to be
regularized, which was done in \cite{Roca:2005nm} either with a
cutoff or in dimensional regularization. In an Effective Theory with
a well defined counting, the
dependence on the regulator could be absorbed into higher
order constants, as is done in ChPT and the IAM. However, when
generating resonances dynamically from the lowest order Lagrangian
 the regularization introduces an
additional parameter in the amplitudes: either a cut-off, or a
subtraction constant for the dimensional regularization case. In
ref. \cite{Roca:2005nm} it was shown that the low lying axial-vector
mesons can be easily generated with a natural cutoff $\Lambda\sim1
\,\gev$.

Although the $N_c$ behavior of the cutoff is not known from QCD, it
is however clear that it cannot grow faster than the cutoff 
of the Effective Theory itself,
 which is of the order of the scale of symmetry breaking
$\Lambda\simeq 4 \pi f_\pi$. Otherwise we would have the absurd
situation that we can extend the validity of the loop integral
beyond the applicability of the theory. Therefore, a natural
integral cutoff, as it is the case here, could scale as
$\sqrt{N_c}$, but no faster.

Of course, we will consider the possibility that the cutoff may
scale slower than $\sqrt{N_c}$, since it would be $O(1)$ if it was given by the mass of heavier $q\bar{q}$ mesons, which cannot be
generated from low-energy two-meson dynamics, and therefore
have been integrated out.
Actually, the unitarization can generate $q\bar{q}$ resonances, like the
vector nonet,
but requires cutoffs at the TeV scale, utterly unnatural, or the explicit values of further
parameters that encode information beyond two-body dynamics
\cite{Truong:1988zp,Guerrero:1998ei,Pelaez:2004xp}.
Of course, if $\Lambda$ starts growing like $\sqrt{N_c}$,
{\it for sufficiently large $N_c$}, it will
reach the
first $q\bar{q}$ meson mass and will behave as $O(1)$.
Hence, we expect a natural behavior in between $O(\sqrt{N_c})$ and $O(1)$.
This will be nicely confirmed later on, with the use of the Weinberg sum rule.

However, a priori, we do not know when the $O(1)$ behavior sets in, because
the lightest heavy state might not have a $q\bar q$ nature and its mass
might not be $O(1)$, also, the width of $q\bar q$ states decreases very fast
as $1/N_c$ and, as $N_c$ grows, their effect is only felt if the energy is
very close to their mass.

Let us now remark that, actually, we are interested in the $N_c$
behavior not far from $N_c=3$, since we study the nature of
the physical states, which, most likely, have a
small admixture of different kinds of ``bare'' or ``preexistent''
states with different $N_c$ behavior. Since the proportions
in this admixture could change with $N_c$, we are not actually
interested in very large $N_c$ since the meson composition at
such large $N_c$ could be completely different from the one
observed at $N_c=3$.
 For instance, since  $q\bar{q}$ states survive as $N_c\rightarrow \infty$, even the tiniest $q\bar{q}$ admixture may show up for a sufficiently large $N_c$. It
is then crucial not to take $N_c\rightarrow \infty$.
In other words, we are {\it only} studying the {\it leading} $N_c$
behavior of a given pole that should be visible not far from
$N_c=3$. Hence, we keep $N_c<20$, i.e., a variation by less
than an order of magnitude.

\begin{figure}[t]
\begin{center}
\includegraphics[scale=0.92,angle=-90]{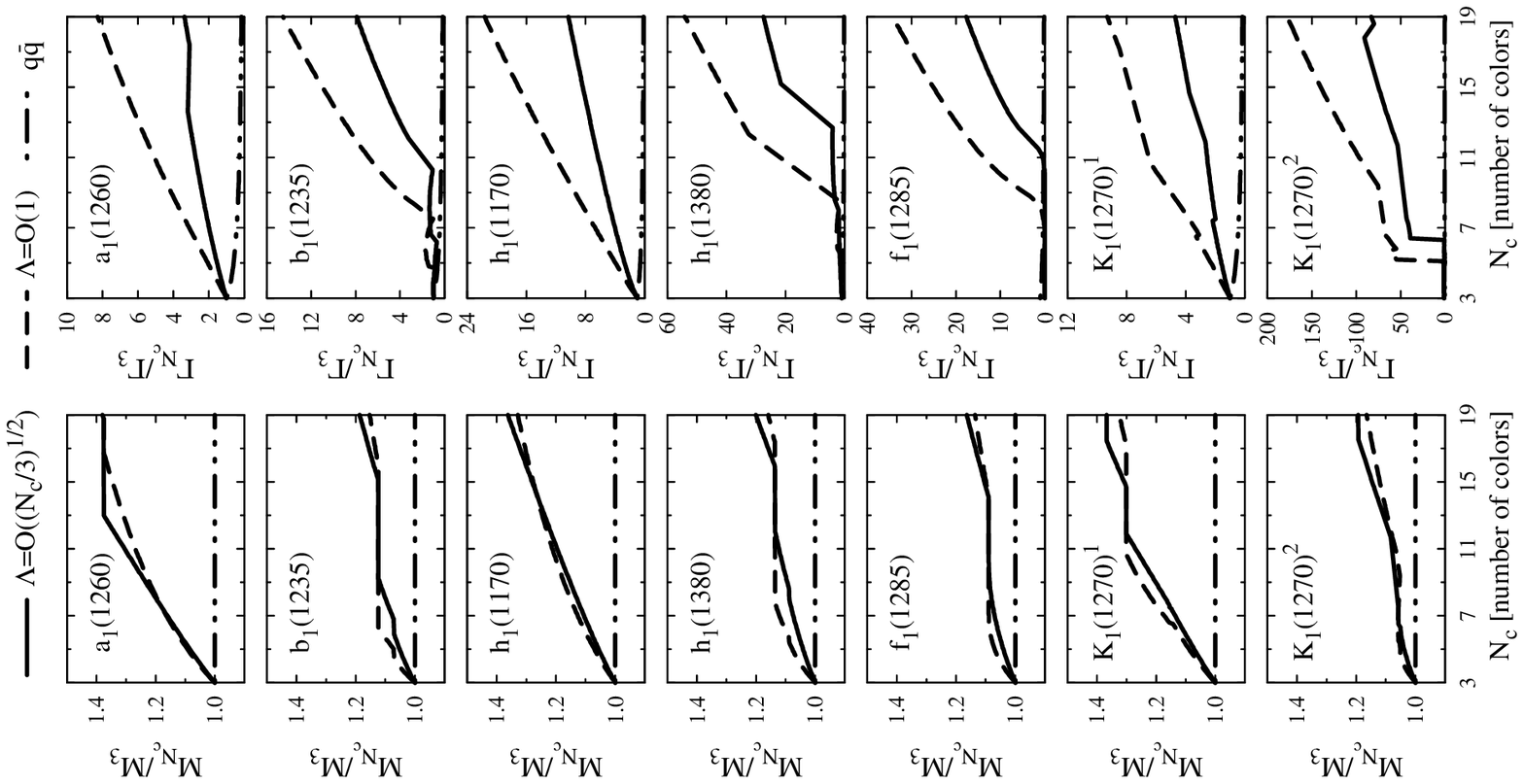}
\caption{Mass and width $N_c$ behavior of the axial-vector mesons.
$K_1(1270)^1$ and $K_1(1270)^2$ denote the low-energy and
high-energy states associated to the nominal $K_1(1270)$. The
$f_1(1285)$ has a zero width at $N_c=3$ in our model, but for
convenience of comparison, we assign it its width of 24.2 MeV in the
PDG~\cite{Yao:2006px} at $N_c=3$. \label{fig1}}
\end{center}
\end{figure}

\begin{figure}[t]
\begin{center}
\includegraphics[scale=.50,angle=-90]{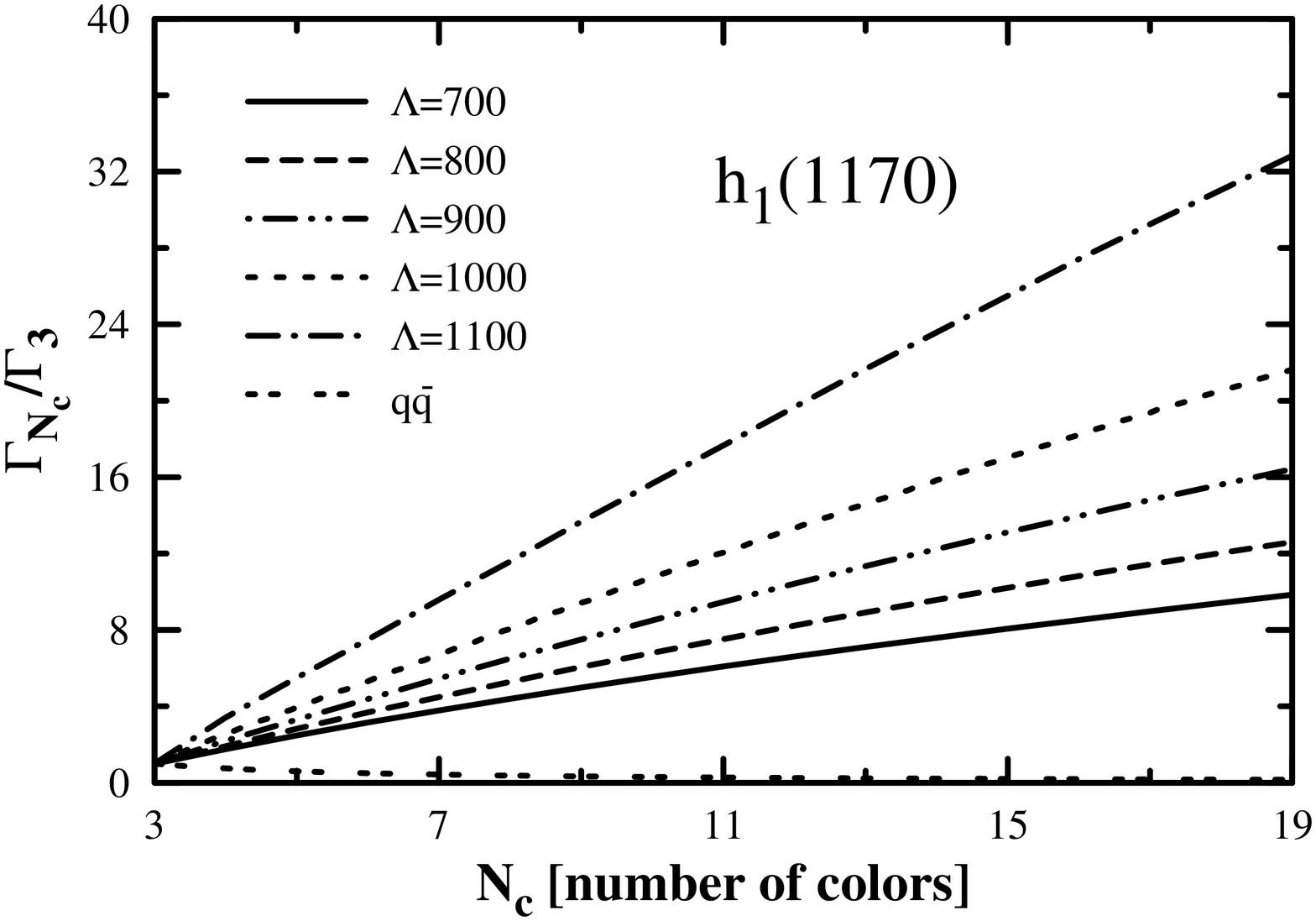}
\caption{\rm $N_c$ behavior of the $h_1(1170)$ width for different
choices of a natural cutoff scaling with $N_c$ as $O(1)$. Note that
in all cases the width grows with $N_c$ in contrast with the $1/N_c$
behavior of $q\bar{q}$ states. Similar results are found
for the other light axial-vector mesons. \label{fig2}}
\end{center}
\end{figure}

Thus, by scaling $f_{\pi N_c}\rightarrow f_\pi \sqrt{N_c/3}$
and $\Lambda$ in the amplitudes of ref.\cite{Roca:2005nm} we obtain the large $N_c$ dependence of the poles associated to the
different axial-vector mesons. Since $N_c$ is not too large,
we cannot guarantee that the $O(1)$ scaling regime is reached
and therefore we study two cases: $\Lambda\rightarrow
\Lambda\, \sqrt{N_c/3}$, which is the largest allowed growth, and
$\Lambda$ constant. The non-trivial fact that
our results do not depend on that choice allows us to make a firm
statement about the axial-vectors nature.

Thus, in Fig.~\ref{fig1}  we show the $N_c$ evolution of the masses and widths of the
axial-vector resonances, normalized to their values at $N_c=3$.  We study
the $a_1(1260)$, $b_1(1235)$, $h_1(1170)$, $h_1(1380)$, $f_1(1285)$ and the
two $K_1(1270)$ resonances. In ref.~\cite{Roca:2005nm} these states were
found from the interaction of several coupled channels in each case and it
was found that they coupled most importantly to the $\rho\pi$, $K^*\bar K$,
$\rho\pi$,  $K^*\bar K$, $K^*\bar K$ and $K^*\pi$, $\rho K$, respectively,
with different quantum numbers.

Both the masses and widths are taken from the associated
pole position, using $\sqrt{s_{pole}}\simeq M-i\Gamma/2$, which is a
very good approximation since all these resonances have $\Gamma<<M$.
For reference, we have plotted as a dash-dotted line the {\it leading} behavior for $q\bar{q}$ states, namely, $M_{N_c}/M_3=
1$ and $\Gamma_{N_c}/\Gamma_3=1/N_c$. From the mass behavior not
much can be concluded, since, although there is an increase in its
value, it is of the order of $30\%\simeq 1/3$ for $N_c\approx20$,
not incompatible with a possible $q\bar{q}$ nature, either with
$\Lambda$ behaving as $\sqrt{N_c}$ or as a constant. This is
nevertheless the order of magnitude of the mass in the case of the
non $q\bar{q}$ states $\sigma$ or $\kappa$ scalar mesons found in
\cite{Pelaez:2003dy}, although in the case of the $\sigma$ there was
a large dispersion of the values.

However, we find that the widths {\it do grow} with $N_c$ in sharp contrast
with the QCD $1/N_c$ behavior for the widths of $q \bar{q}$ states. Let us
remark that this happens irrespective of whether we assume that the cutoff
behaves as $O(1)$ or $\sqrt{N_c}$. Moreover, the slowest growth of the
widths is given for the maximum allowed growth for the natural cutoff
$\Lambda\simeq\sqrt{N_c}$, which means that for the other allowed behaviors
of the cutoff, that grow slower, the $N_c$ dependence of the widths
separates even more from the $q\bar{q}$ behavior. Our result is thus stable
with respect to the different cutoff $N_c$ dependences.

Of course, one could still wonder if our results require
a fine tuning of the cutoff value used to describe the data at $N_c=3$.
Hence, in Fig.~\ref{fig2} we show the $N_c$ dependence of the $h_1(1170)$
width using different cutoffs of natural size. Our previous
arguments are again confirmed to be firm, since in all instances
the width {\it grows} with $N_c$, which is once again against a
dominant $q\bar{q}$ behavior, and the larger the cutoff, the faster
the growth.

In all cases, we find for $N_c$ not larger than 20 a steady increase
of the width of the axial vector mesons. This is in striking
difference from the $1/N_c$ scaling of the width of the $q\bar{q}$
states, followed very clearly by the $\rho$~\cite{Pelaez:2003dy}.
Nevertheless, in Fig.~\ref{fig1} we can see a very different
behavior of the width for different states. In some cases it starts
growing steadily from $N_c=3$ while in other cases the clear growth
appears at larger values of $N_c$. The reason can easily be traced
to the different coupled channels entering the formation of each
state together with the quite different coupling of the state to
these coupled channels~\cite{Roca:2005nm}. As $N_c$ grows, 
the thresholds of the important channels are opened and
the steady growth of the width appears. Just for the sake of
comparison with the scalar mesons of \cite{Pelaez:2003dy} we
effectively parameterize the curves for the $a_1(1260)$ and
$h_1(1170)$ as $a+b(N_c/3)^\alpha$ with the values: $a=-1.91$,
$b=2.79$ and $\alpha=0.70$ for $a_1(1260)$ and $\Lambda\sim 1$;
$a=-0.51$, $b=1.51$ and $\alpha=0.59$ for $a_1(1260)$ and
$\Lambda\sim \sqrt{N_c}$ (up to $N_c=14$); $a=-4.78$, $b=5.72$ and
$\alpha=0.83$ for $h_1(1170)$ and $\Lambda\sim 1$; $a=-2.41$,
$b=3.57$ and $\alpha=0.69$ for $h_1(1170)$ and $\Lambda\sim
\sqrt{N_c}$.
We see an increase of the width slightly
slower than $O(N_c)$ which was also the case for both the $\sigma$
and $\kappa$ in \cite{Pelaez:2003dy} (where $0.5< \alpha< 1$).
However the coefficient in front of $(N_c/3)^\alpha$ is larger than
unity in our case while in \cite{Pelaez:2003dy} was around one,
which indicates a faster growth of the width as a function of $N_c$ in the
present case.

\begin{figure}[t]
\begin{center}
\includegraphics[scale=1.,angle=-90]{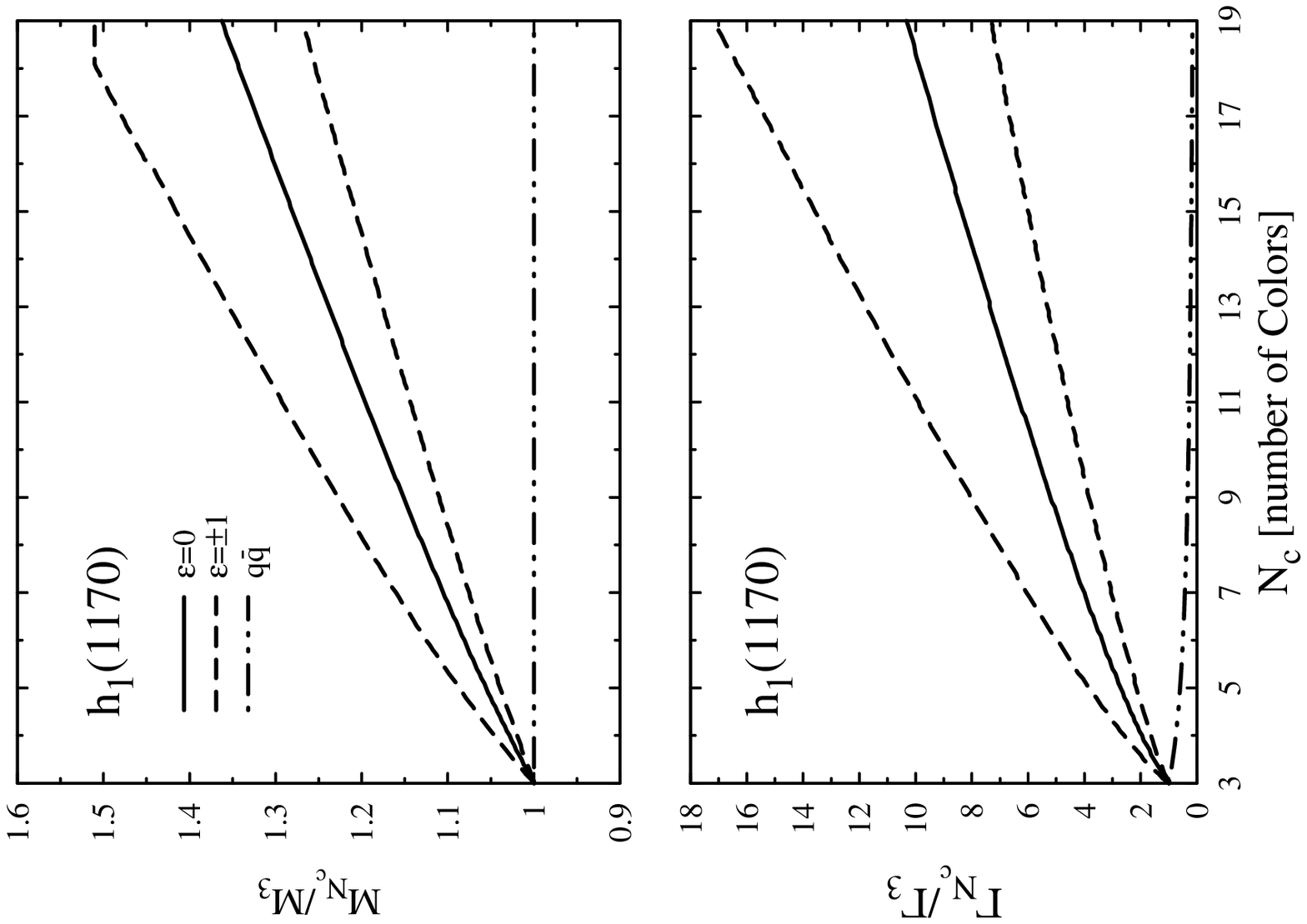}
\caption{\rm The dashed lines cover the estimated uncertainty in the $N_c$ behavior of the $h_1(1170)$ mass and width due to subleading $1/N_c$ contributions
to $f_\pi$. The continuous line assumes $f_\pi$ scaling just as $\sqrt{N_c}$. Even though we show the case with the cutoff
scaling as $\sqrt{N_c}$, the behavior is still completely at odds with a dominant
$q\bar q$ nature (dot-dot-dashed line). Similar results are found for the other resonances.
\label{fig3}}
\end{center}
\end{figure}

On the other hand,
 one could also wonder about the stability of our results on subleading $1/N_c$ corrections to $f_\pi$. Actually, such corrections come partly from the $O(p^4)$ terms
that renormalize the tree level decay constant $f_0$ within standard ChPT \cite{chpt1} (see also the second reference in \cite{Guerrero:1998ei} for a simplified expression). They have also been estimated
within a quark model in \cite{Espriu:1989ff}. For our purpose it
is enough to observe that all these can be recast under the general form:
\begin{equation}
f_{\pi\,N_c}^2= \frac{N_c}{3} \frac{f_\pi^2}{1+\epsilon/3} \,
 (1+\epsilon/N_c)
\label{fsub}
\end{equation}
which ensures the correct physical value at $N_c=3$, the correct leading behavior, but also a subleading contribution of the expected size, namely, $30\%$,
when $\epsilon=1$.
For $\epsilon=0$ we recover our previous calculation.
Actually, this uncertainty roughly covers the subleading terms estimated in \cite{chpt1} and \cite{Espriu:1989ff}.
Thus, in Fig.~\ref{fig3} we show as an example, with the region between the two dashed lines,
the effect of this uncertainty on the $h_1(1170)$ resonance $N_c$ dependence.
Similar results are found for the other generated resonances.
Let us remark that the qualitative behavior is very conclusive,
since, once again,  both the mass and width increase with $N_c$,
in sharp contrast with the behavior
for a $q\bar{q}$ state.
Note that, since the uncertainties due to the $1/N_c$ suppressed contributions 
to $f_\pi$ are similar both for the
$\Lambda\sim \sqrt{N_c}$ and $\Lambda\sim O(1)$ cases, we are not showing the
latter, where the difference with
a $q\bar{q}$ behavior is even more significant.\\

Let us comment now on a possible improvement of the Lagrangian
considered in the present work.
The results shown are evaluated with the kernel
 $V$, Eq.~(\ref{eq:Vij}), of the Bethe-Salpeter equation.
  However, an improvement of this kernel
comes from the addition of the term provided 
by the $SU(3)$ breaking 
Lagrangian \cite{Prades:1993ys,Cirigliano:2006hb}.

\be
{\cal L}=\lambda_m \langle V_\mu V^\mu \chi_+\rangle
\ee
where $\chi_+=u^\dagger\chi u^\dagger+ u \chi^\dagger u$, with 
$\chi=diag(m_\pi^2,m_\pi^2,2m_K^2-m_\pi^2)$ and $u=\exp(iP/\sqrt{2}f)$.

The parameter $\lambda_m$ is readily evaluated form the $K^*$-$\rho$
mass splitting

\be
\lambda_m=\frac{M_{K^*}^2-M_\rho^2}{4(m_K^2-m_\pi^2)}
\ee 
which is of order ${\cal O}(N_c^0)$ in the $1/N_c$ expansion.

Expanding $\chi_+$ up to two pseudoscalar meson fields, one obtains an
interaction term for $VP\to VP$ to be added to that 
of Eq.~(\ref{eq:Vij}). If one looks at the $\rho\pi$ channel the new
term is proportional to $m_\pi^2$ and is negligible compared to   
Eq.~(\ref{eq:Vij}). But this is not the case in channels with
strangeness, where it is proportional to $m_K^2$. For instance, for
$\rho^+ K^-\to \rho^+ K^-$ the new term can be 30\% of the dominant one.
Since the corrections are not negligible, a reanalysis of the results of  
ref.~\cite{Roca:2005nm} with the new term would be of interest, maybe
improving on the semiquantitative agreement with data found in 
ref.~\cite{Roca:2005nm}. However, for the purpose of the present work,
such exercise does not change the conclusions. Indeed, the new potential
is proportional to $\lambda_m m^2/f^2$ and, hence, scales like the one
of Eq.~(\ref{eq:Vij}). But, as we have seen, as $N_c$ grows, so do the
masses of the axial-vectors generated, and hence the variable $s=M_A^2$,
as a consequence of which the strength of the $\lambda_m$ term  becomes
progressively smaller relative to the one of Eq.~(\ref{eq:Vij}), used in
the present work. As an example, for $N_c=19$
and the $\rho^+ K^-\to \rho^+ K^-$ channel, the ratio of terms is
smaller than 10\%.
\\


At this point one might wonder about 
the Weinberg sum rule \cite{Weinberg:1967kj}
that
using chiral symmetry imposes a strong relation between the vector and axial
spectral functions and is expected to hold independently of the number 
of colors. We will show next that this relation is naturally satisfied 
within the uncertainties of our approach, and could even provide a more refined
estimate of the cut-off $N_c$ dependence\footnote{We are indebted to the
referee for this suggestion.}. Given the precision of our
whole approach, the usual estimation of the sum rule 
considering just one pole contribution should be enough. In such case,
the sum rule reads 
\be
F_V^2-F_A^2\simeq f_\pi^2,
\label{eq:WSR}
\ee 
where $F_V$ is the coupling of the 
vector resonance to the vector current and
$F_A$ is the coupling of the axial-vector resonance to the axial current.
We will show that our approach can effectively
generate an axial coupling constant $F_A$ and that the cutoff $N_c$
behavior needed to
satisfy exactly the one-pole estimation of the sum rule lies right in between
the two extreme behaviors we have been considering so far. 

The definition and normalization of the $F_A$ 
and $F_V$ couplings is similar as in the familiar case when 
explicit $A_\mu$ fields are considered with the following interaction terms
\ba
{\cal L}&=&-\frac{F_V}{\sqrt{2}M_V}<\partial_\mu V_\nu f^{\mu\nu}_+>,
\label{eq:LFV}\\
{\cal L}&=&-\frac{F_A}{\sqrt{2}M_A}<\partial_\mu A_\nu f^{\mu\nu}_->,
\label{eq:LFA}
\ea
which are given in an equivalent form in ref.~\cite{Ecker:1988te} with
antisymmetric tensors for the vector meson fields, with
\ba
f^{\mu\nu}_\pm&=&u F_L^{\mu\nu}u^\dagger\pm u^\dagger F_R^{\mu\nu}u,\nn \\
F_L^{\mu\nu}&=&\partial^\mu l^\nu-\partial^\nu l^\mu-i[l^\mu,l^\nu],\nn\\
F_R^{\mu\nu}&=&\partial^\mu r^\nu-\partial^\nu r^\mu-i[r^\mu,r^\nu],\nn\\
r^\mu&=&v^\mu+a^\mu,\qquad
l^\mu=v^\mu-a^\mu.
\ea
Expanding $f^{\mu\nu}_\pm$ up to one pseudoscalar meson field, the
former Lagrangians give us the coupling of the explicit axial-vector resonance,
 $A_\mu$, to an axial-vector current, $a_\mu$, 
through $F_A$ and the coupling of $VP$
to the same current via $F_V$.

\begin{figure}[t]
\begin{center}
\includegraphics[scale=0.4,angle=0]{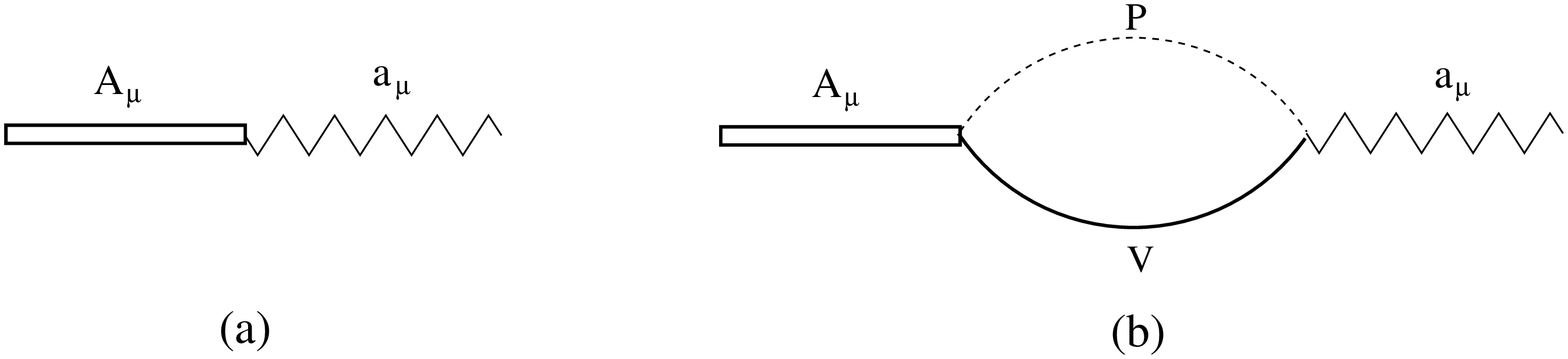}
\caption{\rm 
Coupling of an axial-vector resonance to an external 
axial-vector current via the Lagrangian of Eq.~(\ref{eq:LFA}), (a),
 or in the case of
the resonance being dynamically generated, (b).
\label{fig:Aa}}
\end{center}
\end{figure}

Of course, in our scheme the axial-vector
resonances are dynamically generated, and do not appear explicitly in the Lagrangian, but the resonance is still linked to the
axial-vector current via a $VP$ loop, as depicted 
in Fig.~\ref{fig:Aa}(b).
For the particular case of the $a_1(1260)$ resonance, generated basically 
through $\rho\pi$ loops,
which is largely the dominant channel, from  Fig.~\ref{fig:Aa}(b) we read 
\be
F_A=\frac{F_V}{\sqrt{2}f}\,\frac{M_A}{M_V}\,\frac{s-m_\pi^2+m_\rho^2}{s}\,
 G_{\rho\pi}(s)\,g_{\rho\pi},
\label{eq:FAvsFV}
\ee
where $G_{\rho\pi}(s)$ is the loop function of $\rho\pi$ 
(see Eq.~(\ref{eq:loopVP})) appearing in 
Fig.~\ref{fig:Aa}(b) and $g_{\rho\pi}$ the coupling constant of
the dynamically generated $a_1(1260)$ to $\rho\pi$ in isospin $I=1$,
which is obtained within our approach from the residues of the
$\rho\pi\to\rho\pi$ amplitude at the resonance pole.
In the numerical evaluation of Eq.~(\ref{eq:FAvsFV}), we take
$M_V=M_\rho$ and $M_A=\textrm{Re}(\sqrt{s_{pole}})$. 
Eq.~(\ref{eq:FAvsFV}) provides complex values for the $F_A$ coupling.
Thus, in order to compare with the $F_A$ defined in Eq.~(\ref{eq:LFA}),
which has different phase,
we take for  $F_A$ in the following its absolute value. 

Let us recall that we have shown that, although 
for our calculations we have used a  central value 
$\Lambda(N_c=3)=1000\mev$, we still obtain a 
fair description of data and
the non-$q\bar{q}$ behavior within a much wider range of the cutoff. 
If we now impose the one-pole approximation of the
Weinberg sum rule,  to get exactly $f_\pi=92\mev$ in  Eq.~(\ref{eq:WSR}),
using $s=s_{pole}$ in Eq.~(\ref{eq:FAvsFV}) we find
 $\Lambda(N_c=3)=785\mev$.
This simple one-pole estimate shows that within our approach 
it is fairly simple to accommodate the Weinberg sum rule for 
physical $N_c=3$. Note that, given the precision of our approach,
we can neglect other possible 
resonance and continuum contributions\footnote{JRP thanks S. Leupold for 
comments about this continuum contribution.}.

\begin{figure}[ht]
\begin{center}
\includegraphics[width=0.7\textwidth,angle=0]{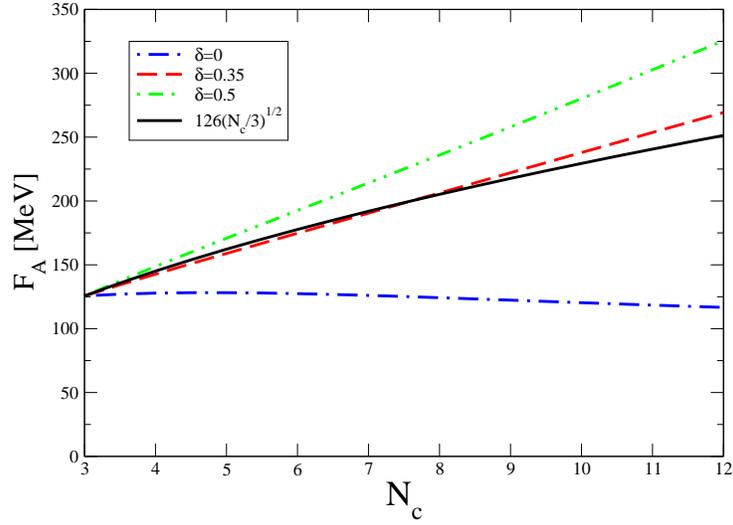}
\caption{Dependence on the number of colors, $N_c$, of the axial-vector
coupling, $F_A$, for different $N_c$ dependences of the cutoff.}
\label{fig:figWSRFA}
\end{center}
\end{figure}

But now we can impose the one-pole approximation of the Weinberg sum rule
for larger $N_c$, thus estimating 
the cutoff $N_c$ dependence, which, for the sake of simplicity, we 
will parameterize as $\Lambda\sim O(N_c^\delta)$. Thus, in
Fig.~\ref{fig:figWSRFA} we plot the resulting $N_c$ dependence of $F_A$ 
from Eq.~(\ref{eq:FAvsFV})
for different $\delta$ values. 
Note that for $\delta\sim0.35$ 
we find the $F_A\simeq F_A(N_c=3)\sqrt{N_c/3}$ behavior.
Of course, this value of $\delta$ is just an estimate, since we are using
the one-pole approximation of the Weinberg sum rule, and we have not allowed
for subleading $1/N_c$ uncertainties in $F_V$, $\Lambda$ or $f_\pi$.
However, it is remarkable and reassuring to note that it comes out naturally
 right in-between the two extreme cases of cutoff behavior
that we had already considered, namely, $\delta=0$ and $\delta=1/2$.
Hence, we have shown that the Weinberg sum rule can be easily accommodated
within our approach and the predominantly non-$q\bar{q}$ 
behavior of the axial-vector resonances.

In summary, we have studied the $N_c$ behavior 
of the axial-vector mesons using the unitarized chiral approach with a
phenomenological Lagrangian for vector and pseudoscalar
mesons, which has been shown to generate and describe the known phenomenology
of axial-vector resonances. This model does not require
any fine-tuning since it has just two parameters: 
the pion decay constant, whose $N_c$ dependence is known from QCD,
and a cutoff of a very natural size. We have shown here that
assuming a natural $N_c$ scaling of the cutoff, either $O(1)$ if
it is due to more massive mesons, or $O(N_c)$ if it is due to the chiral scale,
the resulting $N_c$ behavior of the generated resonances
is not that of predominantly $q\bar q$ states.  In particular, their
widths grow as $N_c$ increases not too far from the physical $N_c=3$
value, contrary to the QCD $N_c$ behavior of $q\bar q$ states. 
Of course, a smaller $q\bar{q}$ component may
not be excluded, but it is not predominant. This growth
is always obtained in our approach and is faster than that found
for other non-predominantly $q\bar{q}$ states, like the light
scalars. This suggests a rather natural interpretation of the axial vector resonances as
non-$q\bar q$ states.

We also estimated the Weinberg sum rule that,
using chiral symmetry, imposes a strong relation between the vector and axial
spectral functions and should hold independently of the number 
of colors. For the physical $N_c=3$ case, and using the 
usual one narrow resonance saturation approximation of the sum rule, 
we have shown that it can be accommodated 
within the natural uncertainties of the cutoff. 
If such approximation of the sum rule is also imposed for larger
 values of $N_c$,
the cutoff behavior lies naturally 
in-between the two extreme cases considered in this work.
Consequently, without the need of any fine tuning,
 the sum rule is found to be consistent with our 
predominantly non-$q\bar{q}$ behavior of axial vectors.

We should also note that the results obtained are not trivial at all if 
we look
at them from the perspective of the findings of 
\cite{Hyodo:2007np,Roca:2008kr} where, in the case of the meson-baryon
interaction generating two $\Lambda(1405)$ states, in the large $N_c$ limit
the pole associated with the singlet becomes a bound state, with zero width
while the other state fades away.

The method we have followed is remarkably simple and could
be easily extended to other systems where dynamically generated
states are obtained within the Chiral Unitary approach with a
natural cutoff and with constants whose leading QCD $N_c$ behavior
is known. We consider that the technique we have presented could provide a method to identify resonance candidates whose spectroscopic nature
may not be predominantly that of a $q\bar q$ state.

\vspace{.3cm}

\textbf{Acknowledgments:} L. S. Geng thanks the Ministerio de Educacion y Ciencia in the Program of
estancias de doctores y tecnologos extranjeros for finantial support. 
 This work is partly
supported by DGICYT contracts FPA2007-29115-E, FIS2006-03438,
FPA2005-02327 and FPA2007-62777,
 Santander/Complutense contract PR27/05-13955-BSCH, 
 the Fundaci\'on S\'eneca
 contract 02975/PI/05, the
Generalitat Valenciana and the EU Integrated
Infrastructure Initiative Hadron Physics Project under contract
RII3-CT-2004-506078.


\end{document}